\documentclass[superscriptaddress,aps,prd,reprint]{revtex4-2}

\usepackage{lipsum}
\usepackage{blindtext}
\usepackage{graphicx}

\usepackage{dcolumn}% Align table columns on decimal point
\usepackage{bm}% bold math

\usepackage{hyperref}
\usepackage{slashed}
\usepackage{amsfonts}
\usepackage{mathrsfs}
\usepackage{color}
\usepackage[makeroom]{cancel}
\usepackage{graphicx}
\usepackage{amssymb}
\usepackage{amsmath}
\usepackage{mathtools}
\usepackage{hyperref}
\usepackage{cleveref}
\usepackage{slashed}
\usepackage{enumitem}

\def\bi{\begin{itemize}}
\def\ei{\end{itemize}}
\def\be{\begin{equation}}
\def\ee{\end{equation}}
\newcommand{\bea}{\begin{eqnarray}}
\newcommand{\eea}{\end{eqnarray}}

\begin{document}

\title{Parity-violating Dark Photon Halos}

\author{Stephon Alexander}
\email{stephon\_alexander@brown.edu}

\author{Lawrence Edmond IV}
\email{lawrence\_edmond@brown.edu}

\author{Cooper Niu}
\email{cooper\_niu@brown.edu}
\affiliation{Department of Physics, Brown University, Providence, RI 02912, USA}

\date{\today}

\begin{abstract}
\noindent
We propose a mechanism for the generation of gravitationally bound dark photon halos during the matter-dominated era. Coupled to an ultralight axion field through a parity-violating Chern-Simons term, dark photons can be produced by the tachyonic instability of axion coherent oscillation. The dark photons with a net helicity lead to a metric vorticity and can generate chiral substructures. For axion masses in the range $10^{-28} \, \mathrm{eV} \lesssim m_a  \lesssim 10^{-22} \, \mathrm{eV}$, the resulting inhomogeneities collapse to form halos with masses spanning $M_{\rm halo} \sim 10^5 \, M_{\odot}$ to $10^{11} \, M_{\odot}$, with halo sizes ranging from $\mathcal{O}(1)$ to $\mathcal{O}(10^{6}) \, \mathrm{pc}$. During halo collapse, the induced vorticity could mediate efficient angular-momentum transport, which enables monolithic collapse and provides primordial seeds for the early formation of supermassive black holes.
\end{abstract}

\maketitle

%%%%%%%%%%%%%%%%%%%%%%%%%%%%%%%%%%%%%%%
\section{Introduction}
%%%%%%%%%%%%%%%%%%%%%%%%%%%%%%%%%%%%%%%
\noindent
The Cold Dark Matter (CDM) paradigm is the bedrock of modern cosmology. While the CDM has achieved remarkable success in explaining the large-scale structure of the universe, it leaves much to be understood about the small-scale distribution, particle properties, and possible non-gravitational interactions of dark matter. After years of efforts to directly detect dark matter (DM), no conclusive evidence of DM particles has yet been found. 

With the increasing precision in astrophysical observations, there is growing interest in developing new probes of dark sector physics in the sky. The gravitational interaction of DM naturally clusters into gravitationally-bound structures, namely \emph{halos}. The characteristics of these DM halos, such as their abundance, density profiles, and internal substructure, are sensitive to the underlying particle physics of DM~\cite{Bechtol:2022koa, Boddy:2022knd, Sengul:2022edu}. Gravitational lensing has emerged as a particularly promising observational probe of DM halos. Gravitational lensing analysis of bullet clusters provides some of the strongest evidence for DM~\cite{Clowe:2006eq, Bradac:2006er, Randall:2008ppe, Cha:2025swk}. It has also yielded constraints on the mass function of dark matter halos down to $\sim 10^7 \, M_\odot$~\cite{Vegetti:2008eg, Gilman:2019vca, Gannon:2025nhr, Metcalf:2002ra, Vegetti:2023mgp}.

Dark photons are a simple and attractive dark matter candidate. They arise naturally as a consequence of extending the Standard Model (SM) of particle physics with an additional dark $\mathrm{U(1)}$ gauge group. The corresponding gauge boson is the dark photon, $A_\mu'$. Under Higgs mechanism or Stueckelberg mechanism, dark photons can acquire a stable mass term~\cite{Caputo:2021eaa}. Dark photons can be efficiently produced via many channels, including gravitational particle production~\cite{Capanelli:2024rlk, Kolb:2020fwh, Graham:2015rva}, cosmic strings~\cite{Kitajima:2022lre, Long:2019lwl}, inflationary fluctuations, and parametric resonance~\cite{Adshead:2023qiw, Zhang:2025pgb, Kitajima:2023pby, Dror:2018pdh, Bastero-Gil:2018uel, Co:2018lka, Agrawal:2018vin}. Since the spin and mass information of dark matter is unknown, the dark photon, as a spin-1 vector boson, could leave a unique imprint distinguishable from other DM candidates. 

Beyond its role as a dark matter candidate, the dark photon provides a natural framework for exploring parity violation in the dark sector. The Chern-Simons coupling between axion and gauge field sources many parity-violating signatures such as $TB$ and $EB$ correlations in the Cosmic Microwave Background (CMB)~\cite{Lue:1998mq, Adshead:2015jza, Pospelov:2008gg}, primordial magnetogenesis~\cite{Fujita:2015iga, Campanelli:2005ye, Adshead:2016iae, Brandenberger:2025gks}, and cosmic birefringence~\cite{Carroll:1989vb, Carroll:1991zs, Nakagawa:2021nme}. A Chern-Simons-like coupling between axion and a dark $U(1)$ gauge field can induce a fundamental asymmetry in the dark sector's evolution. 

In this work, we explore a novel and complementary scenario of dark photon production in the late universe and the formation of chiral dark photon halos. We consider an ultralight axion-like particle coupled to a dark $U(1)$ gauge field via a Chern-Simons interaction~\cite{Agrawal:2022lsp, Blinov:2021axd, Alexander:2004us, Alexander:2009tp}. During the matter-dominated era, coherent oscillations of the axion induce a helicity-dependent instability in the gauge field that amplifies one transverse helicity of the dark photon over another. A key feature of this mechanism is the generation of \textit{chiral} substructure. The helical dark photon field induces vorticity in the stress–energy tensor, sourcing vector metric perturbations and generating a frame-dragging gravitational field~\cite{Amin:2022pzv}. Although vector perturbations dilute with cosmic expansion, their late-time production in this mechanism ensures that they remain dynamically relevant during structure formation. 

Recent observations from the James Webb Space Telescope (JWST) have uncovered a surprisingly large population of massive early galaxies and supermassive black holes (SMBHs), challenging our understanding of the galaxy formation (e.g.~\cite{Naidu_2022, 2023Natur.616..266L, Castellano_2023, 2024Natur.633..318C}). One widely discussed solution to this early massive galaxy puzzle is ``direct collapse'', in which massive black hole seeds form rapidly from the collapse of low-metallicity gas clouds, bypassing fragmentation into standard stellar populations~\cite{Santarelli:2025uck, Haehnelt:1993yy, Madau:2001sc, 2018MNRAS.476.3523L, 2018MNRAS.479.2277A}. This pathway can naturally produce SMBHs with masses of $10^4\!-\!10^6\,M_\odot$ at very high redshifts. In our mechanism, the vorticity induced by chiral dark photon halos can similarly facilitate direct collapse of dark matter halos, potentially accelerating the formation of early massive galaxies.

For axion masses $10^{-28} \, \mathrm{eV} \lesssim m_a \lesssim 10^{-22} \, \mathrm{eV}$ and decay constant $f_a \sim 10^{16} \, \mathrm{GeV}$, the halo mass scales as $M_{\mathrm{halo}} \propto \phi_{*}^2 / m_a$, yielding typical values from $M_{\mathrm{halo}} \sim 10^{5} \, M_{\odot}$ to $10^{11} \, M_{\odot}$, with length scales of order $R \sim 1-10^6 \, \mathrm{pc}$, spanning from the scale of star clusters to galaxy clusters. Because of the net helicity carried by the dark photon, it generates a parity-odd signature in the lensing field, including distortions in image configurations, lensing-induced rotations, and nontrivial correlations in lensing residual maps~\cite{Greco:2025xtt, Seljak:1995ve}.

This letter is organized as follows. In Section~\ref{sec:mechanism}, we present the dynamics of axion-induced gauge field resonance and estimate the conditions under which significant amplification occurs. Section~\ref{sec:halo} derives the halo mass scale and virial velocity. We conclude in Section~\ref{sec:conclusion} with a summary of our findings and implications for early supermassive black hole formation and potential strong lensing signature.

%%%%%%%%%%%%%%%%%%%%%%%%%%%%%%%%%%%%%%%
\section{The Model}
\label{sec:mechanism}
%%%%%%%%%%%%%%%%%%%%%%%%%%%%%%%%%%%%%%%
We consider the model with one axion field interacting with a dark photon field $A_\mu$. The action of the model is given by
\begin{align}
    \begin{aligned}
        \mathcal{S}= \int d^4x &\sqrt{-g} \Bigg(\frac{1}{2} \partial^\mu \phi \partial_\mu \phi -V(\phi)-\frac{1}{4} F^{\mu \nu} F_{\mu \nu} \\
        & + \frac{1}{2}m_{\gamma'}^2 A^\mu A_\mu-\frac{\lambda}{4 f_a} \phi F^{\mu \nu} \tilde{F}_{\mu \nu} \Bigg),
    \end{aligned}
\end{align}
where $F_{\mu \nu}=\partial_\mu A_\nu-\partial_\nu A_\mu$ is the field strength tensor associated with the dark photon, and $\tilde{F}^{\mu \nu} \equiv \frac{1}{2} \epsilon^{\mu \nu \rho \sigma} F_{\rho \sigma}$ denotes its Hodge dual. Here, $f_a$ is the axion decay constant, $m_{\gamma'}$ is the mass of the dark photon, and $\lambda$ is the dimensionless coupling constant between the axion and the dark photon. The Levi-Civita symbol $\epsilon^{\mu \nu \rho \sigma}$ is defined such that $\epsilon^{0123}=+1$ in a flat spacetime background. 

We are interested in the coherent oscillation of axion that leads to efficient production of dark photon during matter-domination epoch. For concreteness, we consider a cosine axion potential trapped in a local minimum until matter-domination epoch~\cite{Kitajima:2023pby}
\begin{equation}
    V(\phi) = m_a^2f_a^2\left(1-\cos \left(\frac{\phi}{f_a}\right)\right) + V_{\rm trap} (\phi),
\end{equation}
where the trap potential
\begin{equation}
    V_{\mathrm{trap}}(\phi)=\frac{1}{2} m_*^2\left(\phi-\phi_*\right)^2 \theta\left(t_*-t\right)
\end{equation}
is a quadratic potential with a Heaviside step function that vanishes after some critical time $t_*$. The axion field is stabilized at $\phi_*$, which later becomes the initial angle for the \textit{trapped misalignment mechanism}~\cite{Higaki:2016yqk, Nakagawa:2020zjr, Jeong:2022kdr, Kitajima:2023pby, Nakagawa:2022wwm}. In particular,~\cite{Kitajima:2023pby} shows that the delayed axion oscillation can lead to efficient dark photon production even during the late universe. 

Given initial amplitude $\phi_*$ of the axion field at the time of trap release, $t=t_*$, what is the constraint on its amplitude such that it does not oversaturate the observed abundance of dark matter today? After release, if the axion starts oscillating with a small initial field value $\phi < f_a$, the axion potential is approximately quadratic $V(\phi) \approx \frac{1}{2} m_a^2 \phi^2$.  The energy density of the oscillating axion field is then
\begin{align}
    \rho_\phi(t)=\frac{1}{2}\dot\phi^2 + \frac{1}{2}m_a^2\phi^2,
\end{align}
and scales as $\rho \propto a^{-3}$ during matter domination. The present-day axion abundance is obtained by evolving the redshifted energy density $\rho_\phi\left(t_*\right)$ to today. The result is captured by the approximate relation~\cite{Hui:2021tkt, Marsh:2015xka, Hui:2016ltb}
\begin{align}
    \Omega_a h^2 \sim 0.1\left(\frac{\phi_*}{10^{17} \, \mathrm{GeV}}\right)^2\left(\frac{m_a}{10^{-22} \, \mathrm{eV}}\right)^{1 / 2}
\end{align}
In our analysis, we assume the dark photon is produced resonantly during the matter-dominated epoch. This occurs within the non-relativistic regime ($m_{\gamma'} \gg H_*$) at the moment of trap release and further on. As energy is transferred directly from the axion-like field to the dark photon, the total dark matter energy budget remains constant, ensuring the expansion history of the universe is unaffected.

In the Friedmann-Lemaitre-Robertson-Walker (FLRW) cosmology, the spacetime is assumed to be spatially flat and undergo homogeneous and isotropic expansion, described by the metric $g^{\mu\nu} = \mathrm{diag} (-1, a^2, a^2, a^2)$ with $a(t)$ as the scale factor. The equations of motion become
\begin{align}
    \ddot{\phi}+3 H \dot{\phi}+\frac{\partial V}{\partial \phi} &= \frac{\lambda}{f_a} \langle \mathbf{E} \cdot \mathbf{B} \rangle,\label{eq:axion_eom}\\
    \ddot{\mathbf{A}} +H \dot{\mathbf{A}}-\frac{1}{a^2}\nabla^2 \mathbf{A} &+ m_{\gamma'}^2 \mathbf{A}- \frac{1}{a}\frac{\lambda}{f_a} \dot\phi\nabla\times \mathbf{A} = 0, \label{eq:gauge_eom}
\end{align}
where $a(t)$ is the scale factor with $H \equiv \dot a/a$ is the Hubble parameter. Under the isotropy assumption, the axion can pick up a background value, while the dark photons are treated as a perturbation of the dark $\mathrm{U(1)}$ gauge field. In our following analysis, we treat the axion as a classical and homogeneous background field and expand the gauge field into Fourier modes,
\begin{align}
    \mathbf{A}(t, \mathbf{x})= \sum_{\lambda= \{\pm, L\}}\int \frac{d^3 k}{(2 \pi)^3}  \bm{\epsilon}_\mathbf{k}^{\lambda} \mathbf{A}_\mathbf{k}^\lambda(t) e^{i \mathbf{k} \cdot \mathbf{x}} + \mathrm{h.c.},
\end{align}
where $\bm{\epsilon}_\mathbf{k}^{\lambda}$ are the polarization vectors for the two transverse ($\pm$) and one longitudinal modes ($L$) and satisfies $\mathbf{k} \cdot \bm{\epsilon}_{\mathbf{k}}^\lambda = 0$. In temporal gauge $A_0 = 0$, the dark photon mode functions after Fourier decomposition become, 
\begin{align}
    &\ddot{A}_k^{\mathrm{L}} +\left(\frac{3 (k/a)^2 + m_{\gamma'}^2}{(k/a)^2+m_{\gamma'}^2}\right) H \dot A_k^{\mathrm{L}} + \left(\frac{k^2}{a^2} + m_{\gamma'}^2\right)A_{\mathrm{L}} = 0 \label{eq:eom_longitudinal}\\
    &\ddot{A}_k^{\pm}+H \dot{A}_k^{\pm} +\left(\frac{k^2}{a^2} + m_{\gamma'}^2 \mp \frac{k}{a} \frac{\lambda}{f_a}\dot{\phi}\right) A_k^{\pm} =0 \label{eq:eom_transverse}.
\end{align}
The longitudinal mode $A_k^{\mathrm{L}}$ evolves under an effective time-dependent friction coefficient 
\begin{align}
    \Gamma_{\rm eff} = \left(\frac{3 (k/a)^2 + m_{\gamma'}^2}{(k/a)^2+m_{\gamma'}^2}\right) H
\end{align}
which interpolates between $3H$ in the radiation-like regime and $H$ in the matter-like limit. In contrast, the transverse modes $A_k^{\pm}$ can experience a tachyonic instability when the axion velocity $\dot\phi$ is sufficiently large due to the Chern-Simons coupling. The unstable modes lie between the finite band, $\xi_{-} < k/a < \xi_+$ where
\begin{equation}
    \xi_{\pm} = \frac{\lambda \dot{\phi}}{2 f_a}\left(1\pm\sqrt{1-\frac{4m_{\gamma'}^2 f_a^2}{\lambda^2 \dot{\phi}^2}}\right).
\end{equation}
Immediately after the trap release at $t=t_*$, axion energy dominates over dark photon, and hence the non-linear term $\langle \mathbf{E}\cdot \mathbf{B}\rangle$ in Eq.~\eqref{eq:axion_eom} is too small to have a substantial effect to the axion field. At that moment, the axion background field takes the form $\phi\left(t\right) = \Phi\left(t\right)\cos\left(m_at\right),$ with $\Phi(t)$ the time-dependent axion oscillation amplitude. Various lattice simulations found the dark photon production halts after the the non-linear backreaction becomes substantial~\cite{Agrawal:2018vin, Co:2018lka, Kitajima:2017peg, Zhang:2025pgb}. We focus on the linear analysis of the dark photon production as our goal is to study the dark photon halo formation.

Assuming the homogeneous oscillation of axion field, we may rewrite Eq.~\eqref{eq:eom_transverse} into a Mathieu equation
\begin{equation}
\label{eq:Mathieu Equation}
    \frac{d^2A_{k}^{\pm}}{dx^2} + \left[F_k \mp q_{k}(x) \sin\left(x\right)\right]A_{k}^{\pm}=0,
\end{equation}
where we define $F_k = (\frac{k}{am_a})^2 + (\frac{m_{\gamma'}}{m_a})^2$ and $q_k = \frac{k}{am_a}\frac{\lambda\Phi(x)}{2f_a}$ for convenience and the effective frequency is $\omega_k^2(x)=F_k-q_k(x) \sin(x)$. According to Floquet theory~\cite{Amin:2014eta, landau2013course}, we can write the solutions of $A_k^\pm$ as 
\begin{align}
    \label{eq:floquet_ansatz}
    A_k^\pm (x) = \Pi_{1,k}(x)e^{-\mu_kx} + \Pi_{2,k}(x)e^{\mu_kx},
\end{align}
where $\Pi_1(x)$ and $\Pi_2(x)$ are periodic functions and $\mu$ is the Floquet exponent. Modes with non-zero $\mathrm{Re}[\mu_k]$ admit exponential growth solutions. Notice that the amplitude of the axion background field is time-dependent. Since the density scales as $\rho\sim a^{-3}$ during the matter-dominated epoch, we find that the amplitude of the axion scales as $\Phi(t) \sim a^{-3/2}$ during this period. When the system is evolving slowly, one can approximate the mode over one cycle using Wentzel–Kramers–Brillouin (WKB) method, 
\begin{align}
    A_k^\pm (x) \sim \exp\left( \int^{x} \mu_k(x') \, dx' \right)
\end{align}
If the system evolves adiabatically, the vacuum state deforms smoothly, and a state with zero particles stays a state with zero particles. However, if the change is non-adiabatic, the old vacuum state becomes a state populated with particles in the new configuration. Specifically, the WKB (adiabatic) condition for a mode is given by $|\dot{\omega}_k/\omega_k^{2}| \ll 1$. Such condition is violated if $\omega_k \to 0$. We therefore use WKB only for analytic intuition, and compute the exact instability bands numerically.
\begin{figure}[h]
    \centering
    \includegraphics[width=0.9\linewidth]{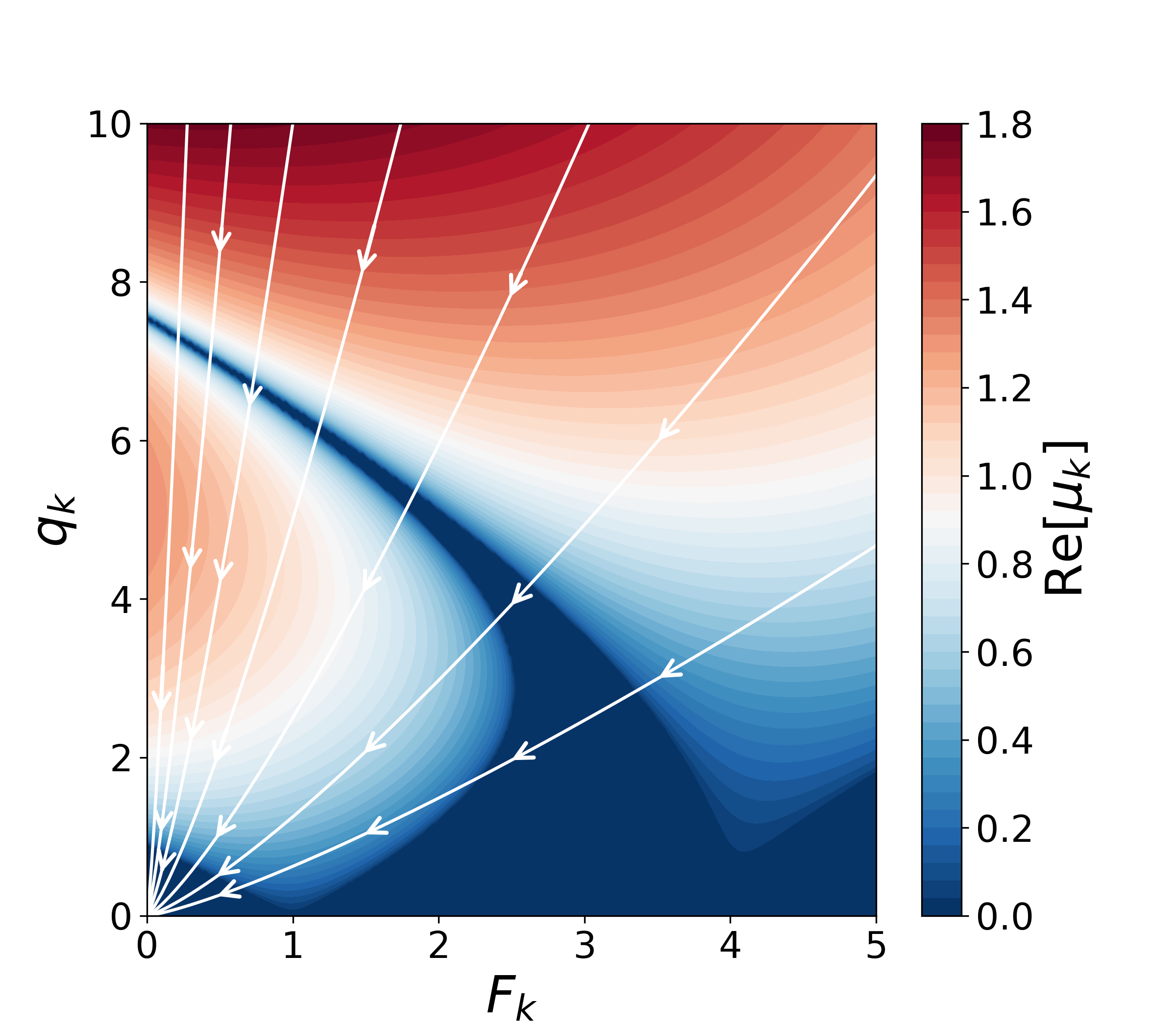}
    \caption{Stability diagram of the Mathieu equation. The color scale represents the Floquet exponent, representing the growth rate of unstable modes. White trajectories show the evolution of individual Fourier modes, with arrows indicating the direction of time flow.}
    \label{fig:floquet stability}
\end{figure}

Fig.~\ref{fig:floquet stability} illustrate the real part of Floquet exponent $\mathrm{Re}[\mu_k]$ mapped across the $(F_k,q_k)$ parameter space. The primary resonance peak, visible as the most prominent instability tongue emerging at $k \sim am_a$, represents the most efficient band for particle production. Within these shaded regions, the real part of the Floquet exponent is positive, indicating that the dark photon field modes grow exponentially as $A_k^\pm \propto e^{\mu_k x}$. Along with the cosmic expansion, the resonance bands shift, causing the comoving mode $k$ (white lines) to sweep across multiplet resonance bands. Physically, this generates a broad-spectrum population of dark photons. 
\begin{figure*}
    \centering
    \includegraphics[width=0.48\textwidth]{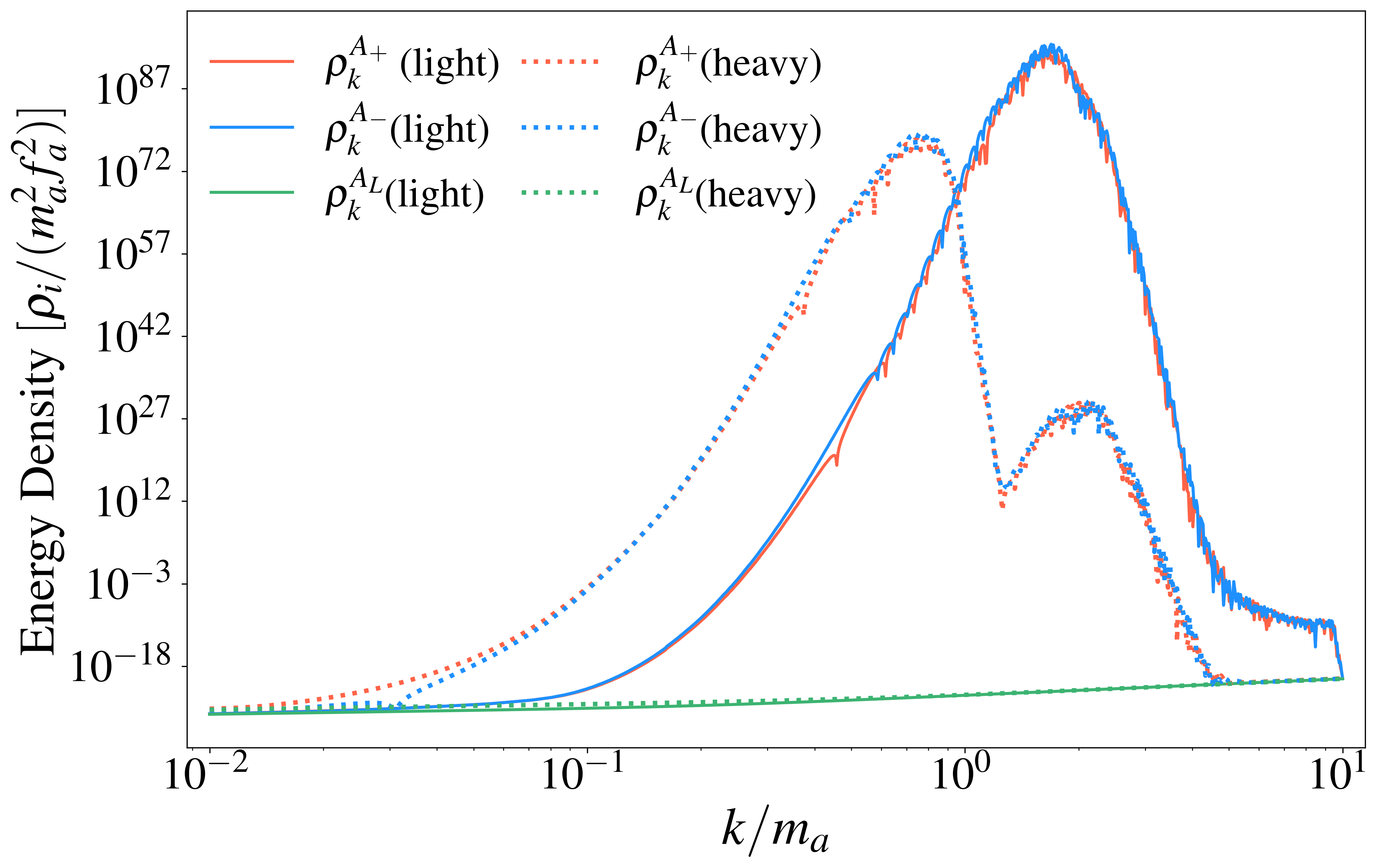}
    \hfill
    \includegraphics[width=0.46\textwidth]{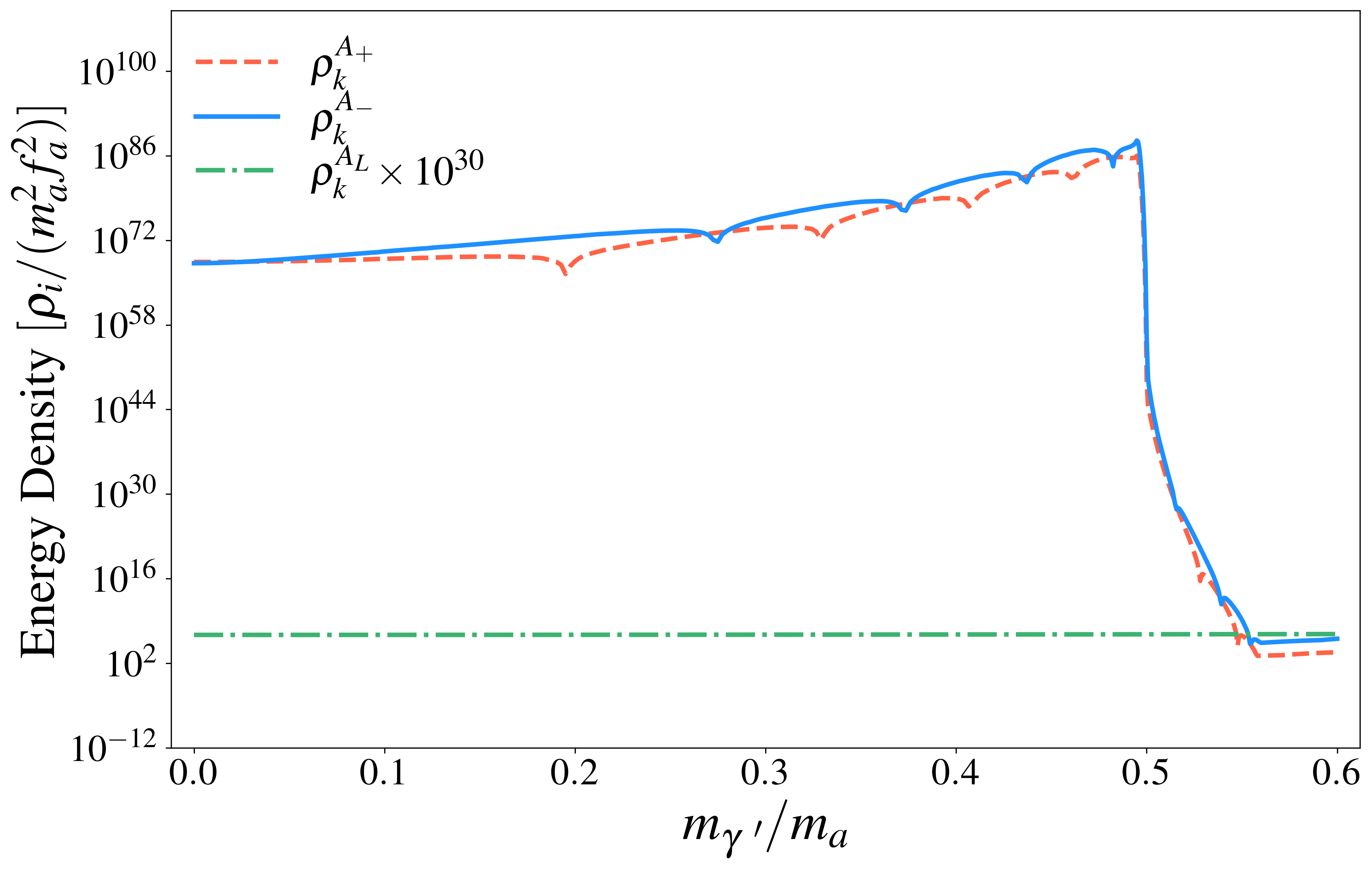}
    \caption{\textbf{Left:}  Power spectrum of produced dark photons from axion background oscillations. The blue and red solid lines represent the transverse polarizations for the heavy ($m_{\gamma'} = 0.5 m_a$) and light ($m_{\gamma'} = 0.2 m_a$) dark photon cases, respectively. \textbf{Right:} the energy density of dark photon polarization modes as a function of axion-dark-photon mass ratio $m_{\gamma'}/m_a$. The longitudinal mode is scaled by $10^{30}$ for visibility. For both plots, we assume $m_a = 10^{-29}~\text{eV}$, $\lambda = 5$, $f_a = 10^{17}~\text{GeV}$, $z_* = 20$, and the initial axion angle $\phi_*/f_a = 1$.}
    \label{fig:evolution}
\end{figure*}
We numerically solve the axion-dark-photon system. Since our goal is to have a matter-domination vector perturbation generation mechanism not realistic simulation, we assume the backreaction $\langle \mathbf{E}\cdot\mathbf{B}\rangle$ is negligible for simplicity. Moreover, we assume the initial dark photon abundance comes only from the quantum fluctuation in primordial universe and adopt Bunch-Davies initial condition.

We present the dark photon energy power spectrum in  Fig.~\ref{fig:evolution} (left). For the heavy dark photon ($m_{\gamma'} = 0.5 m_a$), the spectrum is dominated by a single, large-amplitude peak, indicating a strong primary resonance. In contrast, the light dark photon ($m_{\gamma'} = 0.2 m_a$) exhibits a multi-peaked structure with two smaller, distinct maxima, which arises as the lower mass allows the comoving modes to traverse multiple instability bands. Both transverse modes experience tachyonic instability, but one mode tends to dominate another depending on the initial angle of trapped axion. The longitudinal mode is not amplified, as it does not couple to the homogeneous axion field linearly.   

Fig.~\ref{fig:evolution} (right) shows the final energy density of the dark photon polarization modes as a function of the mass ratio $m_{\gamma'}/m_a$. The transverse modes, $A_k^+$ and $A_k^-$, exhibit a strong dependence on the mass ratio, with a dominant resonance peak occurring as $m_{\gamma'}/m_a$ approaches $0.5$. This result is consistent with~\cite{Agrawal:2018vin}. Conversely, in the relativistic limit ($m_{\gamma'} \to 0$), the resonance condition becomes largely independent of the dark photon mass, and the amplification is instead dominated by the comoving momentum $k$.

%%%%%%%%%%%%%%%%%%%%%%%%%%%%%%%%%%%%%%%
\section{Dark Photon Halo}
\label{sec:halo}
%%%%%%%%%%%%%%%%%%%%%%%%%%%%%%%%%%%%%%%
We now estimate the physical characteristics of the dark matter halos seeded by the resonantly amplified gauge field modes. For our benchmark analysis, we adopt an ultralight axion mass $m_a \sim 10^{-22} \, \text{eV}$ and a decay constant $f_a \sim 10^{17} \, \text{GeV}$, typical of Fuzzy Dark Matter (FDM) candidates. As demonstrated in Fig.~\ref{fig:evolution} (left), the primary peak of the power spectrum occurs near the dimensionless wavenumber $k/m_a \sim 1$. This corresponds to a comoving wavelength
\begin{align}
    \lambda_{\mathrm{comoving}} = \frac{2\pi}{k} \sim \frac{2\pi}{a_* m_a}
\end{align}
Here, $a_*$ is the scale factor at the time of resonance. At the halo turnaround, the comoving radius $R$ of the nascent collapsing region scales as
\begin{align}
    R = 4\times \left(\frac{1+z_*}{20}\right)\left(\frac{10^{-22} \, \mathrm{eV}}{m_a}\right) \, \mathrm{pc}
\end{align}
The energy density of the background axion field is given by $\rho_\phi \simeq \frac{1}{2} m_a^2 \phi_*^2$, where $\phi_*$ is the axion field amplitude at the onset of the resonance. We model the energy transfer efficiency by a parameter $\epsilon \lesssim 0.1$, representing the fraction of axion potential energy converted into the gauge field energy density
\begin{equation}
    \rho_A \approx \frac{\epsilon}{2} m_a^2 \phi_*^2 
\end{equation}
Assuming the energy transfer is efficient and quasi-instantaneous at the scale of turnaround, the total mass $M_{\text{halo}}$ contained within the collapsing volume $V = \frac{4\pi}{3} R^3$ (using the physical radius at formation) can be estimated. Substituting the relationship for $R$, we find
\begin{widetext}
\begin{align}
    M_{\rm halo} \sim 2.7\times 10^{5} \,M_{\odot} \left( \frac{\epsilon}{0.1} \right)\left(\frac{1+z_*}{20}\right)^3 \left(\frac{\phi_{*}/f_a}{1}\right)^2 \left( \frac{10^{-22} \, \mathrm{eV}}{m_a} \right) \left(\frac{f_a}{10^{16} \, \mathrm{GeV}}\right)^2.
\end{align}
\end{widetext}
This estimate places the resulting object in the low-mass mini-halo regime. If the axion mass is lighter, this mechanism can form Dwarf-galaxy halos. As illustrated in Fig.~\ref{fig:evolution} (right), the efficient dark photon production requires $m_{\gamma'} \lesssim m_a/2$. At such tiny masses, the de Broglie wavelength of the dark photon is of astrophysical scale. During halo collapse, wave interference and gravitational cooling cause the inner region to lose excess energy and relax into a stable, localized, and ground-states field configuration called \textit{soliton}. These dark photon solitonic cores can be extremely polarized and are capable of carrying a large angular momentum, which could significantly alter the tidal stripping profiles and the internal kinematics of the halo~\cite{Zhang:2024bjo, Jain:2021pnk, Amin:2022pzv, Chen:2024vgh, Gorghetto:2022sue, PhysRevD.111.043031}. We leave a detailed treatment of solitonic core formation for future works.

The virial velocity serves as an additional diagnostic for the gravitational binding and long-term survivability of these halos against tidal stripping or feedback. Using the standard expression
\begin{align}
    v_{{vir}} \sim \sqrt{\frac{G M_{\mathrm{halo}}}{R}}
\end{align}
where $R \sim \lambda_{\mathrm{phys}} = 2\pi/m_a$. 
Plugging in our previous expressions, we obtain
\begin{align}
    v_{\rm vir} \sim 38 \, \mathrm{km/s} \left( \frac{\epsilon}{0.1} \right)^{1/2}\left(\frac{1+z_*}{20}\right) \left( \frac{\phi_{*}/f_a}{1} \right) \left(\frac{f_a}{10^{16} \, \mathrm{GeV}}\right).
\end{align}
Interestingly, the axion mass $m_a$ cancels out in this expression. This indicates that the gravitational potential well is universal across different axion masses. 

The presence of a chiral dark gauge sector opens several avenues for observational signatures.  In particular, the intrinsic helicity of the dark photons induces both gradient- and curl-type deflections in photon geodesics, resulting in odd-parity gravitational lensing of the CMB temperature and polarization anisotropies~\cite{Hwang:2025ipp, Cooray:2005hm}. Additionally, the energy-momentum density of these fields introduces anisotropic stress and a non-zero Poynting vector ($\delta T^{0}{}_{i} \sim \mathbf{E}\times\mathbf{B}$), generating vector perturbations in the spacetime background. These vortices can seed rotational modes in collapsing overdensities and subtly alter the dynamics of gravitational collapse. Intriguingly, the dark-sector Poynting flux can mediate angular-momentum transport, enabling efficient halo collapse and accelerating structure formation.

%%%%%%%%%%%%%%%%%%%%%%%%%%%%%%%%%%%%%%%
\section{Discussion and Conclusion}
\label{sec:conclusion}
%%%%%%%%%%%%%%%%%%%%%%%%%%%%%%%%%%%%%%%
In this letter, we explore parity-violating structure formation during the matter-dominated epoch, focusing on the late-time amplification of a dark \rm U(1) gauge field. We consider a trapped misalignment mechanism, in which an ultra-light axion field couples to the dark photon via a Chern–Simons term, driving a resonant, Floquet-type instability that amplifies the transverse helicities of the gauge field asymmetrically. This mechanism naturally generates intrinsically helical structures in the dark sector, chiral dark photon halos, with $M_{\rm halo} \sim 10^{5} M_\odot - 10^{11} M_\odot$ for axion masses in the range $10^{-28}\,{\rm eV} \lesssim m_a \lesssim 10^{-22}\,{\rm eV}$, and an axion decay constant \(f_a \sim 10^{16}\,{\rm GeV}\). As vector perturbations are not sourced by the standard $\Lambda$CDM cosmology, this mechanism, as a sustained source of rotational vorticity during matter-domination, may alter the cosmic history of structure formation. 

An implication of this framework is the potential role of these dark chiral halos as \emph{primordial seeds} for the supermassive black holes~\cite{Naidu_2022, 2023Natur.616..266L, Castellano_2023, 2024Natur.633..318C} (SMBHs) observed by the \textit{James Webb Space Telescope} (JWST) at $z \gtrsim 7$. In conventional scenarios, forming $10^9 M_\odot$ black holes by such early epochs requires either (i) massive initial seeds ($10^{4-6} M_\odot$) at $z \gtrsim 15$~\cite{Koushiappas:2003zn, Jiao:2025kpn} or (ii) sustained super-Eddington accretion~\cite{Bogdan:2023ilu}. Our mechanism naturally satisfies the first condition by generating a population of massive, high-density dark halos. Some of the halos could form black hole seeds with masses $M_{\rm seed} \sim 10^{4} - 10^{6}\,M_\odot$ at redshifts $z \gtrsim 15$. Such early seeds would have ample time to grow into the $10^{9} M_\odot$ SMBHs observed by JWST under standard or mildly super-Eddington accretion histories. 

For our halos to undergo gravitational collapse without fragmenting, their temperature (or velocity dispersion) must remain sufficiently high to support monolithic collapse~\cite{Jiao:2025kpn}. This can be understood by comparing the halo’s cooling and free-fall timescales: if the cooling time exceeds the free-fall time, the fluid resists clumping, preventing the formation of gravitationally bound inhomogeneities. A warm halo maintains sufficient internal kinetic energy and sound speed, yielding a large Jeans mass. If the halo reaches a critical mass that allows collapse while preserving this thermal support, it can undergo a relatively homogeneous, non-fragmented collapse. Specifically, the Jeans mass is given by \cite{1994sse..book.....K, 2008gady.book.....B}
\begin{align}
    M_J \sim \frac{\pi^{5/2} c_s^3}{6 G^{3/2} \rho^{1/2}},
\end{align}
where $c_s$ is the sound speed, and the free-fall timescale is given by 
\begin{align}
    t_{\rm ff} \approx \sqrt{\frac{3 \pi}{32 G \rho}}.
\end{align}
At $z_* \sim 20$, $\rho_{\mathrm{halo}} \sim 4.2 \times 10^3 \,~M_\odot \,\mathrm{pc}^{-3}$ with associated Jeans mass and free-fall timescale, $M_{\rm J} \sim 160 c_{\rm s}^3 ~M_\odot \left(\mathrm{km/s}\right)^{-3}$ and $t_{\rm ff}\sim 0.13 \,~\mathrm{Myr}$. For our halo mass $M_{\mathrm{halo}}\sim 2.7\times10^5 \,~M_\odot$, there exists a sub-virial velocity regime in which $M_{\mathrm{halo}} \gtrsim M_{\rm J}$, where the halo is globally Jeans-unstable while remaining warm enough to disfavor small-scale fragmentation. A rigorous demonstration of direct collapse in our proposed mechanism requires a detailed analysis of the cooling processes in the halo, as well as an analysis of the angular momentum transport mediated by the dark gauge fields. However, our discussion shows that such a scenario is plausible in our framework. 

The helical gauge stresses act analogously to magnetic braking: by transferring angular momentum outward, they allow the inner regions of the halos to undergo near-radial collapse. If the dark sector provides an efficient ``cooling'' channel via the Poynting flux $\mathbf{E}\times\mathbf{B}$, gravitational binding energy can be dissipated without fragmenting the structure. In this regime, the central densities can reach conditions conducive to direct collapse into black holes, either entirely within the dark sector or by subsequently accreting baryons through gravitational coupling~\cite{Aggarwal:2025pit, Jiao:2025kpn}.

These results motivate dedicated magnetohydrodynamic and $N$-body simulations that incorporate parity-violating gauge sectors to track the nonlinear evolution of these dark halos, their angular-momentum transport, and possible black-hole collapse pathways. Future surveys of CMB polarization would provide indirect tests of this scenario. 

\section{Acknowledgments}
We thank Mustafa Amin, Robert Brandenberger, Evan Mcdonough, and Savvas Koushiappas for important comments and discussion on an early draft. S.A., L.E., and C.N. acknowledge support from the Simons Foundation, Award 896696.

\bibliographystyle{apsrev4-2}
\bibliography{ref}

\end{document}